\documentclass[ALICE,manyauthors]{cernphprep}
\pdfoutput=1
\usepackage{hyperref}
\usepackage[comma,square,numbers,sort&compress]{natbib}
\usepackage{float}
\usepackage{amstext}
\usepackage{graphicx}
\usepackage{lineno}
\usepackage{array}
\usepackage[nodayofweek]{datetime}
\usepackage[Symbolsmallscale]{upgreek}

\begin{document}%

%%%%%%%%%%%%%%%  Title page %%%%%%%%%%%%%%%%%%%%%%%%
\begin{titlepage}
\PHyear{2017}
\PHnumber{255}      % required, will be obtained from PH
\PHdate{22 September}  % required, will be obtained from PH
%

%%% Put your own title + short title here:
\title{Production of deuterons, tritons, $^{3}$He nuclei and their antinuclei
in pp collisions at $\mathbf{\sqrt{{\textit s}}}$~=~0.9, 2.76, and 7~TeV}

\ShortTitle{Production of light nuclei and antinuclei in pp collisions}   % appears on right page headers

%%% Do not change the next lines
\Collaboration{ALICE Collaboration\thanks{See the Appendix~\ref{app:collab} for the list of collaboration members}}
\ShortAuthor{ALICE Collaboration} % appears on left page headers, do not change

\begin{abstract}
Invariant differential yields of deuterons and antideuterons
in pp collisions at $\sqrt{s}$ = 0.9, 2.76 and 7~TeV and
the yields of tritons, $^{3}$He
nuclei and their antinuclei at $\sqrt{s}$ = 7~TeV have been measured
with the ALICE detector at the CERN Large Hadron Collider. The measurements
cover a wide transverse momentum ($p_{\text{T}}$) range in
the rapidity interval $|y|<0.5$, extending both the energy and the
$p_{\text{T}}$ reach of previous measurements up to 3~GeV/$c$ for $A=2$ and
6~GeV/$c$ for $A=3$. The coalescence parameters
of (anti)deuterons and $^{3}\overline{\text{He}}$ nuclei exhibit
an increasing trend with $p_{\text{T}}$ and are found to be compatible with
measurements in pA collisions at low $p_{\text{T}}$ and lower energies.
The integrated yields decrease by a factor of about 1000 for each increase
of the mass number with one (anti)nucleon. Furthermore, the deuteron-to-proton
ratio is reported as a function of the average charged particle multiplicity 
at different center-of-mass energies.
\end{abstract}
\end{titlepage}
\setcounter{page}{2}

\section{Introduction}

The production of light nuclei and antinuclei has been measured in many 
experiments at energies lower than those of the Large Hadron Collider (LHC).
Deuterons and antideuterons are copiously produced in heavy-ion collisions 
\citep{Bevalac81,E814,E886,NA44,E864,E864-dbar,NA49-d,E896,NA49,PHENIX2005,ALICE-dHe3-PbPb},
but less abundantly in lighter particle collisions, such as pp \citep{ISR,ISR-2} 
and $\overline{\text{p}}$p \citep{Alexopoulos2000} collisions, photo-production 
$\upgamma$p \citep{H1} and e$^{+}$e$^{-}$ annihilation at
$\Upsilon(nS)$ \citep{CLEO-2007} and Z$^{0}$ \citep{ALEPH} energies.
Measurements of heavier antinuclei, such as antitritons and $^{3}\overline{\text{He}}$ 
nuclei, have only been reported in pA \citep{Antipov70,Vishnevskii74}
and AA collisions \citep{Appelquist96,STAR2001,NA52-2003,Agakishiev2011,ALICE-dHe3-PbPb}.

The high luminosity provided by the LHC allows these measurements to be 
extended to higher energies and transverse momenta ($p_{\text{T}}$) than 
in previous experiments, and provides in addition the possibility to detect 
for the first time antinuclei heavier than antideuterons in pp collisions.
Many of these measurements have been explained as the result of the coalescence 
of protons and neutrons that are nearby in space and have similar velocities 
\citep{Zupancic63,Gutbrod76}, but this has not been experimentally tested 
in high $p_{\text{T}}$ regimes in small systems. On the other hand, 
statistical hadronization models~\citep{Andronic2011,ALICE-dHe3-PbPb}
have been successful in describing particle yields over a wide range of energies
in AA collisions, with the chemical freeze-out temperature and baryochemical
potential being constrained by measurements of particle ratios. In this sense, 
the deuteron-to-proton ratio could serve as a test for possible 
thermal-statistical behavior in pp collisions at LHC energies.

On a broader level, this subject may also have an impact on cosmology.
Big-bang nucleosynthesis is the dominant natural source of deuterons
\citep{Wagoner73} and, in the absence of baryogenesis, one could assume
that the same holds for antideuterons. These antinuclei and even heavier
antinuclei can also be produced in pp and pA collisions in interstellar
space, representing a background source in the searches for segregated
primordial antimatter and dark matter \citep{Kounine2012,Mognet2014,Donato2000}.
As it turns out, the low momentum characteristic yields of antinuclei at central
rapidities (compared to forward) lie in an energy region which is best suited
for identification by most satellite-borne (low magnetic-field) instruments,
such as AMS-02 \citep{Kounine2012}.

While the differential yields of deuterons in pp collisions at
$\sqrt{s}$ = 7~TeV have been reported in \citep{ALICE-dHe3-PbPb},
this paper complements the previously published results by providing
the corresponding measurements of antideuterons at the same collision energy.
In addition, results for (anti)deuterons at $\sqrt{s}$ = 0.9 and 2.76~TeV
as well as for (anti)tritons and $^{3}$He (anti)nuclei at $\sqrt{s}$ = 7~TeV
are given. The paper is organized as follows: Section~\ref{sec:experimental-apparatus}
gives a description of the experimental apparatus. Section~\ref{sec:data-analysis}
describes the analysis procedure of the experimental data along with the estimation
of the systematic uncertainties. In Sec.~\ref{sec:results}, the distributions
of (anti)deuterons, (anti)tritons, and $^{3}$He (anti)nuclei are presented.
The integrated yields, the deuteron-to-proton ratios, and the coalescence parameters,
which relate the production of nuclei with those of the nucleons,
are obtained in Sec.~\ref{sec:discussion} and the summary and conclusions
are presented in Sec.~\ref{sec:summary-conclusions}.
 % introduction
\section{Experimental apparatus\label{sec:experimental-apparatus}}

ALICE \citep{ALICE-PPR-VOL1,ALICE-PPR-VOL2,ALICE2008} is a multipurpose
detector designed to study heavy-ion collisions at the LHC and it
also has excellent capabilities to study light nuclei and antinuclei
in pp collisions. The nuclei were identified using the central detectors:
the inner tracking system (ITS), the time projection chamber (TPC)
and the time of flight (TOF) detector. These detectors are located
inside a solenoidal magnetic field with a strength of 0.5~T and cover
the full azimuthal acceptance and the pseudo-rapidity range $|\eta|<0.9$. 

The ITS \citep{ITS2010} consists of six cylindrical layers of position-sensitive
detectors, covering the central rapidity region for vertices located
in $|z|<$10 cm, where $z$ is the distance along the particle beam
direction. The two innermost layers are silicon pixel detectors (SPD),
followed by two layers of silicon drift detectors (SDD), while the
two outermost layers are double-sided silicon strip detectors (SSD).
The ITS is mainly used for reconstruction of the primary and secondary
vertices. It also helps to separate primary nuclei from secondary
nuclei via the determination of the distance of closest approach of
the track to the primary vertex. The TPC \citep{Alme2010}, the main
tracking component of ALICE, is a large drift detector with a low
material budget to reduce multiple scattering and secondary particle
production. In combination with the ITS, it is used to measure particle
momenta. The TPC is also used to identify particles via their specific ionization
energy loss with a resolution of 5\% in pp collisions \citep{Abelev:2014ffa}. The TOF \citep{TOF2013}
detector is a large-area array of multigap resistive plate chambers
covering the full azimuth $0\leq\phi<2\pi$ and 
$|\eta|<0.9$, except the region $260^{\circ}<\phi<320^{\circ}$
and $|\eta|<0.12$ to avoid covering the photon spectrometer with
more material. In pp collisions, it measures the time of flight of particles
with an overall resolution of about 120~ps, allowing the identification of light
nuclei and antinuclei with transverse momenta above 3~GeV/$c$, depending on the
available data. The start time for the time of flight
is provided by the T0 detector, with a time resolution of $\sim40$~ps. 
The T0 consists of two arrays of Cherenkov counters,
T0A and T0C, placed on opposite sides of the interaction point at
$z=375.0$ cm and $z=-72.7$ cm, respectively. If there is no T0 signal,
the TOF detector is used to measure the start time when at least three
particles reach the TOF \citep{Adam:2016ilk} detector.

Between the TPC and TOF detector there is a transition radiation detector (TRD) \citep{ALICE2008}
to discriminate between electrons and pions above 1~GeV/$c$. Only 7 modules out of
18 were installed for the pp run of 2010, leaving the major part of space between
TPC and TOF free of additional material. 
The V0 detector \citep{VZERO2013}, two hodoscopes of 32 scintillator cells each
which cover the pseudo-rapidity ranges $2.8<\eta<5.1$ and $-3.7<\eta<-1.7$,
provides in combination with the SPD the trigger for inelastic pp collisions.

 % experimental apparatus
\section{Data analysis\label{sec:data-analysis}}

The pp events used in this paper were collected by the ALICE Collaboration
during 2010 and 2011. The recorded integrated luminosity for each
analyzed sample is 0.124~nb$^{-1}$, 0.692~nb$^{-1}$, and 4.20~nb$^{-1}$
for the center-of-mass energies $\sqrt{s}$ = 0.9, 2.76, and 7~TeV, respectively.

\subsection{Event and track selection}

The pp events were triggered by requiring a hit in both sides of the
V0, i.e., two charged particles separated by approximately 4.5 units
of pseudorapidity, which suppresses single diffractive events. The
presence of passing bunches was detected by two beam-pickup counters. 
Contamination from beam-induced background was rejected offline
using the timing information of the V0.
Additionally, a cut on the correlation between the number of SPD clusters and 
the number of small track segments (tracklets) in the SPD detector was applied.
Furthermore, in order to maintain a uniform acceptance and to reduce beam-induced
noise, collision vertices were required to be within 10 cm of the
center of the detector in the beam direction and within 1~cm in the
transverse direction. Pile-up events were reduced by requiring that
more than three tracklets or tracks contribute to the reconstructed
vertex. In cases of multiple vertices which are separated by more than
0.8~cm, the vertex reconstruction with the SPD allows these events to be tagged
as pile-up and hence not considered in the analysis.
The events analyzed here consist mostly of nonsingle
diffractive events, which represent a fraction of the total inelastic
cross section equal to $0.763_{-0.008}^{+0.022}$, $0.760_{-0.028}^{+0.052}$, and 
$0.742_{-0.020}^{+0.050}$ for $\sqrt{s}$ = 0.9, 2.76, and 7~TeV \citep{ALICE-Sinel},
respectively. Those fractions were used to extrapolate the measurements
to inelastic pp collisions assuming that the production of nuclei in 
single diffractive events is not significant with respect to nonsingle 
diffractive events based on Monte Carlo estimates (less than 3\%).

For each track at least two track points were required in the ITS
and 70 out of a maximum of 159 in the TPC. A pseudorapidity
cut of $|\eta|<0.8$ was also required to avoid edge effects. 
Tracks with kinks, typically originating from weak decays inside the TPC volume,
were treated as two separate tracks and only the track pointing to the
primary vertex was kept. 

The measurements are reported for the rapidity
interval $|y|<0.5$ and have been corrected for detector efficiency 
based on the GEANT3 particle propagation code \citep{GEANT3}.
Track matching between the TPC and TOF detectors in GEANT3 was 
further improved by a data driven method based on a study of tracks not 
crossing the TRD material, resulting in a 6\% difference.
Since at low $p_{\text{T}}$ many nuclei in $|y|<0.5$ are outside $|\eta|<0.8$,
their number was extrapolated using a Monte Carlo simulation where the
rapidity distribution was approximated by a flat distribution.

In order to allow for a consistent comparison of the antideuteron-to-deuteron
ratio across different center-of-mass energies with an identical GEANT version,
a reanalysis of the deuteron differential yield at $\sqrt{s}$ = 7~TeV is presented here.
The results are found to be consistent with the previous measurements shown
in \citep{ALICE-dHe3-PbPb} within the systematic uncertainties.

\subsection{(Anti)nuclei identification \label{sub:pid}}

\begin{figure}[H]
\noindent \begin{centering}
\includegraphics[scale=0.6]{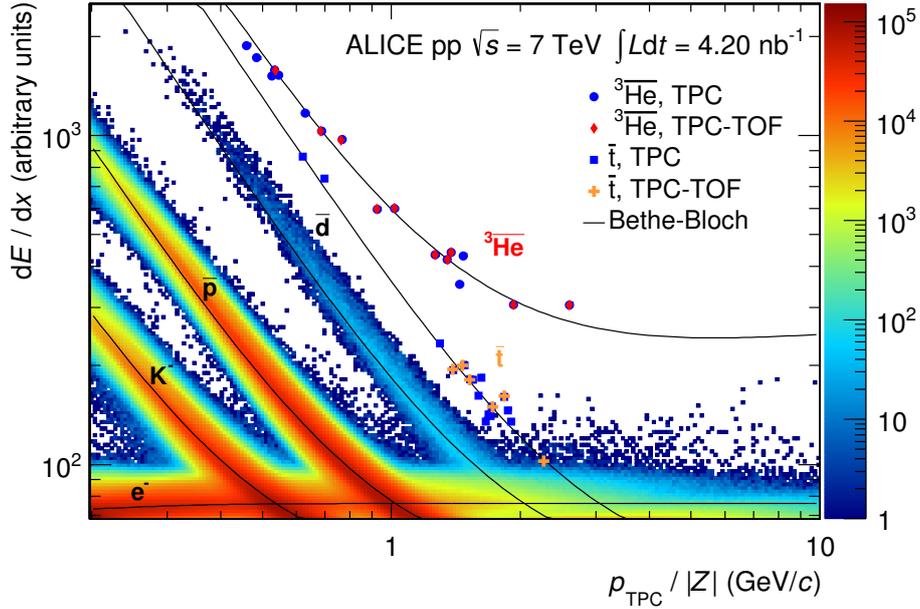}
\par\end{centering}

\caption{\label{fig:dEdx} Energy loss in the TPC ($\text{d}E/\text{d}x)$
of particles with negative charge versus the rigidity estimated at
the TPC inner wall ($p_{\text{TPC}}/|Z|$). The solid lines represent
the expected energy loss according to the parametrization of the
Bethe-Bloch formula. The blue circles and squares are $^{3}\overline{\text{He}}$
nuclei and antitritons identified by the TPC only, and the orange
crosses and the red diamonds are the antitritons and $^{3}\overline{\text{He}}$
nuclei, respectively, that were matched to a TOF detector hit.}
\end{figure}

The identification of nuclei and antinuclei is based on their specific energy
loss in the TPC and the estimation of their mass with the TOF detector.
Figure \ref{fig:dEdx} shows the energy loss signal recorded by the TPC
of different nucleus species versus the rigidity ($p_{\text{TPC}}/|Z|$), where
$p_{\text{TPC}}$ is the momentum estimated at the inner wall of the TPC.
 
\begin{figure}[H]
\noindent \begin{centering}
\includegraphics[scale=0.6]{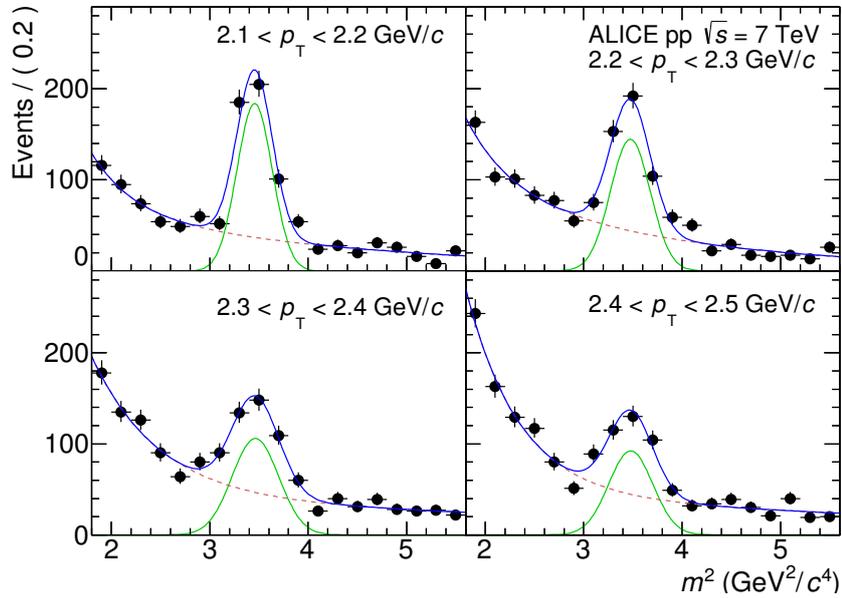}
\par\end{centering}

\caption{\label{fig:DbarDid}Squared mass distribution for tracks within $\pm3\sigma$
of the expected energy loss for antideuterons in the TPC in several
$p_{\text{T}}$ bins. The solid blue line is the global fit, the dashed
line the background, and the green line the antideuteron signal.}
\end{figure}

Deuterons and antideuterons can be identified cleanly
up to $p_{\text{TPC}}\simeq$1.2~GeV/$c$, which corresponds to a maximum 
$p_{\text{T}}$ of 1~GeV/$c$. For $p_{\text{T}}>1$~GeV/$c$ a coincidence with
a TOF signal was required, in addition to a $\pm3\sigma$ cut around their
expected energy loss in the TPC, extending the identification up to
$p_{\text{T}}=3$~GeV/$c$. For this, tracks were propagated to the
outer radius of the TOF detector and, if a hit was found close enough to the
trajectory, the corresponding time of flight was assigned to
the track. Then, the squared mass $m^{2}=p^{2}(t^{2}/l^{2}-1)$ was calculated,
where $p$ is the reconstructed momentum, $t$ the time of flight, and $l$
the track length.
Figure \ref{fig:DbarDid} shows the squared mass distribution 
for several $p_{\text{T}}$ bins in the region of the antideuteron squared mass. 
The antideuteron signal is approximately Gaussian, centered at the deuteron 
squared mass and with an exponential tail on the high mass side. This exponential
tail is also present in the signal of other particle species such as $\uppi$, K,
and p and extends to the antideuteron squared mass, producing an exponential
background. The signal was extracted by combining a Gaussian with an exponential
tail and an exponential background (Fig.~\ref{fig:DbarDid}).

\begin{figure}[H]
\noindent \begin{centering}
\includegraphics[scale=0.6]{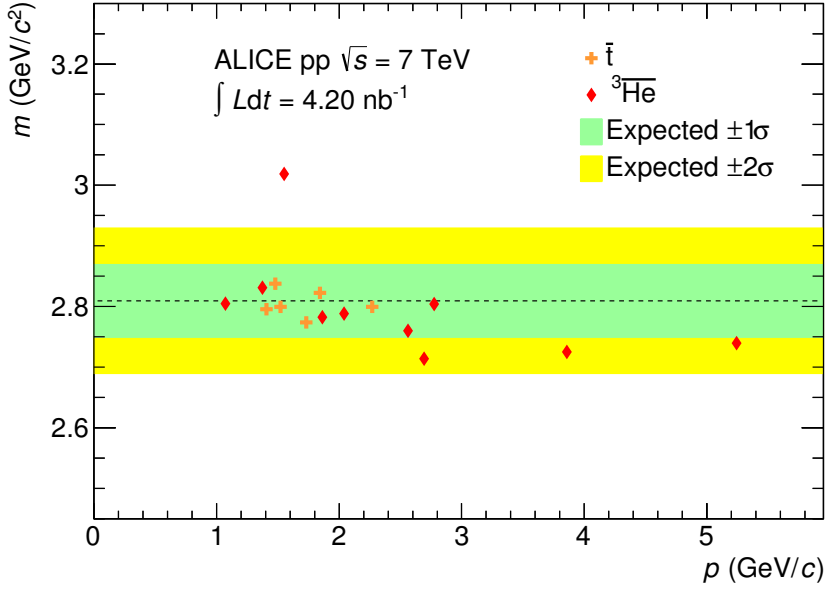}
\par\end{centering}

\caption{\label{fig:He3Mass} Mass distribution of antitriton (crosses) and
$^{3}\overline{\text{He}}$ nucleus (diamonds) candidates obtained
with the TOF detector as a function of the total momentum. The green and yellow
bands represent 1$\sigma$ and 2$\sigma$ intervals, respectively, around the expected $^{3}$He mass
(dashed line), obtained from the TOF resolution.}
\end{figure}

Tritons and antitritons were identified by selecting tracks within
$\pm3\sigma$ of their expected energy loss in the TPC and by also requiring
a match to a TOF detector hit. The minimum $p_{\text{T}}=1.2$~GeV/$c$ was
chosen to be the same as for the $^{3}$He nuclei. Due to the small number
of tritons, it was not possible to use the signal extraction procedure
used for deuterons. In this case, the selected tracks were required to have 
an associated mass within $\pm3\sigma$ ($\sigma\simeq0.05$~GeV/$c^{2}$)
of the triton mass and the maximum $p_{\text{T}}$ was limited to
1.8~GeV/$c$. The result is shown in Figs. \ref{fig:dEdx} and
\ref{fig:He3Mass}, with six antitriton candidates in the interval
$1.2<p_{\text{T}}<1.8$~GeV/$c$.

Unlike deuterons and tritons, $^{3}$He and $^{3}\text{\ensuremath{\overline{\text{He}}}}$
nuclei can be identified throughout the $p_{\text{T}}$ range with
the TPC, since for nuclei with $|Z|=2$ the energy deposition is well
separated from particles with $|Z|=1$. In total, 17 candidates for $^{3}\overline{\text{He}}$
nuclei were observed, based on the specific energy loss in the TPC (Fig.~\ref{fig:dEdx}),
out of which 14 candidates were in the interval $1.2<p_{\text{T}}<6$~GeV/$c$,
and these were used in the measurements.
Their identities were confirmed for those particles that were matched to
a TOF hit (10 out of 14) with a mass measurement based on their times of flight,
as shown in Fig.~\ref{fig:He3Mass}. A few $^{3}\overline{\text{He}}$
nuclei (six candidates) were also observed at the center-of-mass energy 2.76~TeV.

\subsection{Secondary (anti)nuclei \label{sub:secondary-deuterons}}

Secondary nuclei are copiously produced in spallation reactions induced
in the detector material by the impact of primary particles. They
are also produced in the decays of $\Lambda$ hypernuclei, with the
$\uppi$-mesonic decay of the (anti)hypertriton being the most important
contribution \citep{Kamada98}.

\begin{figure}[H]
\noindent \begin{centering}
\includegraphics[scale=0.7]{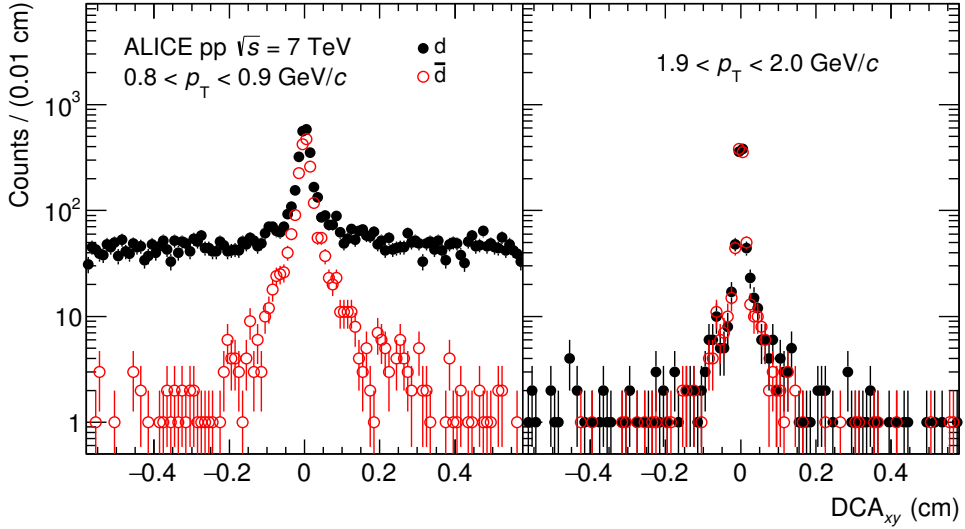}
\par\end{centering}

\caption{\label{fig:DbarDdca}Distance of closest approach in the transverse
plane to the primary vertex (DCA$_{xy}$) of deuterons 
(solid circles) and antideuterons (hollow circles) for the $p_{\text{T}}$ bins
$0.8<p_{\text{T}}<0.9$~GeV/$c$ (left) and $1.9<p_{\text{T}}<2.0$~GeV/$c$
(right). A large background distribution of secondary deuterons is
clearly visible in the left panel.}
\end{figure}

The distances of closest approach (DCA) of the track to the primary
vertex in the transverse plane (DCA$_{xy}$) and along the beam direction
(DCA$_z$) were used to distinguish and reduce the number of secondary nuclei.
Since they are produced far away from the primary
vertex, they have a broader and flatter DCA$_{xy}$ distribution than
primary nuclei, which have a narrow DCA$_{xy}$ distribution peaked
at zero, similar to antinuclei. Figure \ref{fig:DbarDdca} illustrates
the different DCA$_{xy}$ distributions for deuterons and
antideuterons at low and high $p_{\text{T}}$. A positive
DCA$_{xy}$ was assigned when the primary vertex was inside the radius
of curvature of the track and a negative value in the opposite case.
The number of secondary nuclei was greatly reduced by requiring $|\text{DCA}_{xy}|<0.2$~cm
and $|\text{DCA}_{z}|<3$~cm, corresponding to a cut of $\pm10\sigma$ in 
the measured DCA resolution in the lowest $p_{\text{T}}$ bin. 

The fraction of secondary nuclei with respect to primary nuclei
was estimated with DCA$_{xy}$ templates from Monte
Carlo simulations for each $p_{\text{T}}$ bin. The templates were
fitted to the measured distribution with a maximum likelihood procedure
which relies on a Poisson distribution and takes into account both
the measured distribution and Monte Carlo statistical uncertainties
\citep{Barlow93}. This fraction was found to fall exponentially as a 
function of $p_{\text{T}}$ and was subtracted from the measurements.

The production of antinuclei from interactions of primary particles
with the detector material was neglected, since antinuclei exhibit
a narrow DCA$_{xy}$ distribution peaked at zero (Fig.~\ref{fig:DbarDdca}).
Due to the small production cross section of (anti)hypernuclei
in pp collisions, the feed-down contribution of (anti)nuclei 
was not subtracted, but instead included as a systematic uncertainty.

\subsection{Systematic uncertainties}

Table \ref{tab:systematics} summarizes the values of the systematic
uncertainties for the lowest and highest $p_{\text{T}}$ bins. 
These uncertainties take into account the identification procedure,
the track selection criteria, secondary nuclei originating in the detector
material and from feed-down, the (anti)nucleus--nucleus interactions
simulated in GEANT, and the material budget.

The identification procedure was affected by an uncertainty coming
from the background and signal shapes at high $p_{\text{T}}$,
where the signal-to-background ratio was small. It was evaluated by
changing the squared mass interval and extracting the signal with two
different methods: one by using the procedure described in Sec.~\ref{sub:pid}
and the other by counting the number of entries in the $1<p_{\text{T}}<1.4$~GeV/$c$
interval where the identification is unambiguous. For antitritons
and $^{3}\overline{\text{He}}$ nuclei the identification was clean
and the particle identification uncertainty was assumed to be negligible.
Systematic uncertainties due to the track selection criteria were estimated
to be less than 4\% for nuclei and antinuclei by changing the conditions
for selecting tracks. 

The approximations made in the DCA$_{xy}$ templates introduced an uncertainty
on the removal of secondary nuclei originating in the detector material.
A value of 4\% was estimated for deuterons by replacing the simulated DCA$_{xy}$
templates of primary deuterons with the measured DCA$_{xy}$ distribution 
of antideuterons, which are not affected by contamination from secondary tracks.
An uncertainty of $\sim$20\% was estimated following
a similar procedure for tritons and $^{3}$He nuclei.

The dominant feed-down contribution for (anti)nuclei is
the $\uppi$-mesonic decay of (anti)hypertritons \citep{Kamada98}:
$_{\Lambda}^{3}\text{H}\rightarrow\text{d}+\text{p}+\uppi^{-}$,
$_{\Lambda}^{3}\text{H}\rightarrow\text{d}+\text{n}+\uppi^{0}$,
$_{\Lambda}^{3}\text{H}\rightarrow\text{t}+\uppi^{0}$ and
$_{\Lambda}^{3}\text{H}\rightarrow{}^{3}\text{He}+\uppi^{-}$.
In pp collisions, the $_{\Lambda}^{3}\text{H}$ cross section was estimated
to be of the same order of magnitude as the $^{3}$He nucleus cross section
\citep{ALICE-PUBLIC-2017-010}. However, the production cross section of deuterons is about
10$^4$ times greater than that of $^{3}$He nuclei, hence the contamination for (anti)deuterons can be
considered negligible. Additionally,
the fraction of hypertritons which passes the track selection in the
$^{3}$He (anti)nucleus channel was estimated with a Monte Carlo
simulation and is at most 35\%. Assuming a similar value for the
(anti)triton channel and branching ratios of 27\% and 13\% \citep{Kamada98},
then less than $\sim$10\% and $\sim$6\% contamination is expected
for $^{3}$He (anti)nuclei and (anti)tritons, respectively.

The (anti)nucleus--nucleus elastic and inelastic scattering uncertainty
was evaluated by comparing the GEANT3 simulations with the data for 
two different experimental configurations: one using the detector portion in which
the TRD modules were installed between the TPC and the TOF detector
and another in which the TRD was not installed. The ratio between
the number of (anti)deuterons counted with the two different detector configurations is related
to the number of (anti)deuterons lost due to hadronic interactions.
These ratios were compared with a GEANT3 simulation and a 6\% uncertainty
was estimated for the amount of nuclei lost in such processes.
This comparison, however, was not feasible for
(anti)tritons due to the limited data and a 12\% uncertainty was assumed.
Unlike deuterons and tritons, the measurements of $^{3}$He (anti)nuclei presented
here only rely on TPC information, hence they are not affected by the TRD material
in front of the TOF detector.

Another source of systematic uncertainty is the accuracy in the knowledge
of the material budget. This uncertainty was estimated to be $+3.4$\%
and $-6.2$\% by comparing the material thickness estimated by analyzing 
photon conversions in the inner detectors with the material description
implemented in the Monte Carlo simulations \citep{ALICE-pi0}. To propagate these uncertainties
to the results, a Monte Carlo simulation was done in which the material density
was varied by $\pm$10\% and linearly interpolated
to match the uncertainties in the material budget. The result was
below 3\% at low $p_{\text{T}}$ and negligible at
high $p_{\text{T}}$ for the different (anti)nuclei.

The extrapolation of the measurements to inelastic
pp collisions adds additional systematic uncertainties 
of $_{-0.8}^{+2.2}$\%, $_{-2.8}^{+5.2}$\% and $_{-2.0}^{+5.0}$\% 
for the center-of-mass energies 0.9, 2.76,
and 7~TeV, respectively \citep{ALICE-Sinel}. However, these uncertainties
are not included in the figures as in previous related publications 
\citep{ALICE-piKp-900GeV,ALICE_pikp276,ALICE-piKp-7TeV,ALICE-dHe3-PbPb}.

\renewcommand{\arraystretch}{1.2}

\begin{table}[H]
\begin{centering}
\begin{tabular}{lcccccc}
\hline 
 & {\small{}d} & {\small{}$\overline{\text{d}}$} & {\small{}t} & {\small{}$\overline{\text{t}}$} & {\small{}$^{3}$He} & {\small{}$^{3}\overline{\text{He}}$}\tabularnewline
\hline 
{\small{}$p_{\text{T}}$ (GeV/$c$)} & {\small{}0.8 -- 3.0} & {\small{}0.8 -- 3.0} & {\small{}1.2 -- 1.8} & {\small{}1.2 -- 1.8} & {\small{}1.2 -- 3.0} & {\small{}1.2 -- 6.0}\tabularnewline
\hline 
{\small{}Particle identification} & {\small{}negl. -- 20\%} & {\small{}negl. -- 20\% } & {\small{}negl.} & {\small{}negl.} & {\small{}negl.} & {\small{}negl.}\tabularnewline
{\small{}Track selection} & {\small{}4\%} & {\small{}4\%} & {\small{}4\%} & {\small{}4\%} & {\small{}4\%} & {\small{}4\%}\tabularnewline
{\small{}Secondary nuclei} & {\small{}4\%} & {\small{}negl.} & {\small{}18\%} & {\small{}negl.} & {\small{}20\% -- negl.} & {\small{}negl.}\tabularnewline
{\small{}Feed-down nuclei} & {\small{}negl.} & {\small{}negl.} & {\small{}--6\%} & {\small{}--6\%} & {\small{}--10\%} & {\small{}--10\%}\tabularnewline
{\small{}Hadronic interactions} & {\small{}6\%} & {\small{}6\%} & {\small{}12\%} & {\small{}12\%} & {\small{}6\%} & {\small{}6\%}\tabularnewline
{\small{}Material budget} & {\small{}3\% -- negl.} & {\small{}3\% -- negl.} & {\small{}3\%} & {\small{}3\%} & {\small{}2\% -- 1\%} & {\small{}2\% -- 1\%}\tabularnewline
\hline
\end{tabular}
\par\end{centering}

\caption{\label{tab:systematics}Summary of the main sources of systematic
uncertainties for the lowest and highest $p_{\text{T}}$ bins. Symmetric uncertainties
are listed without sign for clarity.}
\end{table}

 % data analysis
\section{Results\label{sec:results}}

\subsection{Deuterons and antideuterons \label{sub:d-and-dbar}}

\begin{figure}[H]
\noindent \begin{centering}
\includegraphics[scale=0.7]{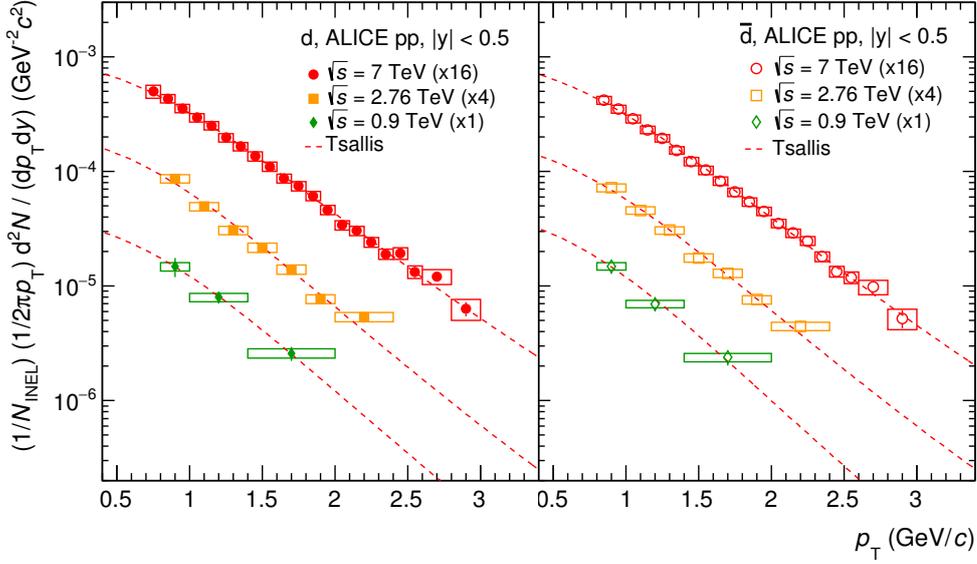}
\par\end{centering}

\caption{\label{fig:DbarDSpectra}Invariant differential yield of deuterons
(left panel) and antideuterons (right panel) in inelastic pp collisions (INEL)
at $\sqrt{s}$ = 0.9, 2.76, and 7~TeV. Systematic uncertainties are represented
by boxes and the data are multiplied by constant factors for clarity in the figure.
The lowest $p_{\text{T}}$ point for deuterons at $\sqrt{s}$ = 7~TeV was taken
from \citep{ALICE-dHe3-PbPb}. The dashed line represents the result of a fit
with a Tsallis function (see Sec.~\ref{sub:yields-mean-pt} for details). }
\end{figure}

\begin{figure}[H]
\noindent \begin{centering}
\includegraphics[scale=0.6]{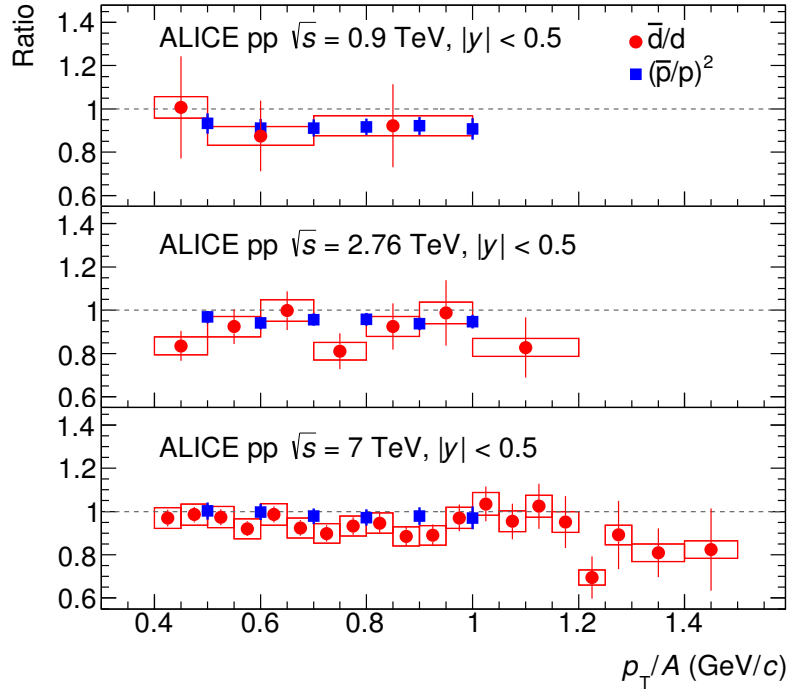}
\par\end{centering}

\caption{\label{fig:DbarDRatio} Antideuteron-to-deuteron ratio ($\overline{\text{d}}/\text{d}$)
as a function of $p_{\text{T}}$ per nucleon in pp collisions compared
with the $(\overline{\text{p}}/\text{p})^{2}$ ratio (squares) at
mid-rapidity ($|y|<0.5$) \citep{ALICE-pbarp,ALICE-BBbar2013}. Boxes represent
the systematic uncertainties and error bars in the $(\overline{\text{p}}/\text{p})^{2}$ ratios
are statistical and systematic uncertainties added in quadrature. }
\end{figure}

The invariant differential yields of deuterons and antideuterons
were measured in the $p_{\text{T}}$ range $0.8<p_{\text{T}}<3$~GeV/$c$ (Fig.~\ref{fig:DbarDSpectra})
and extrapolated to inelastic pp collisions with the cross sections
of Ref.~\citep{ALICE-Sinel}. At LHC energies, both nucleus species are produced
with similar abundance since the antideuteron-to-deuteron ratio approaches
1 as the center-of-mass energy increases (Fig.~\ref{fig:DbarDRatio}).
The ratios are consistent with the $(\overline{\text{p}}/\text{p})^{2}$
ratios extracted from Refs.~\citep{ALICE-pbarp,ALICE-BBbar2013},
and hence are in agreement with the expectation from simple coalescence
and thermal--statistical models.

\subsection{Heavier nuclei and antinuclei \label{sub:tbar-he3bar}}

\begin{figure}[H]
\noindent \begin{centering}
\includegraphics[scale=0.7]{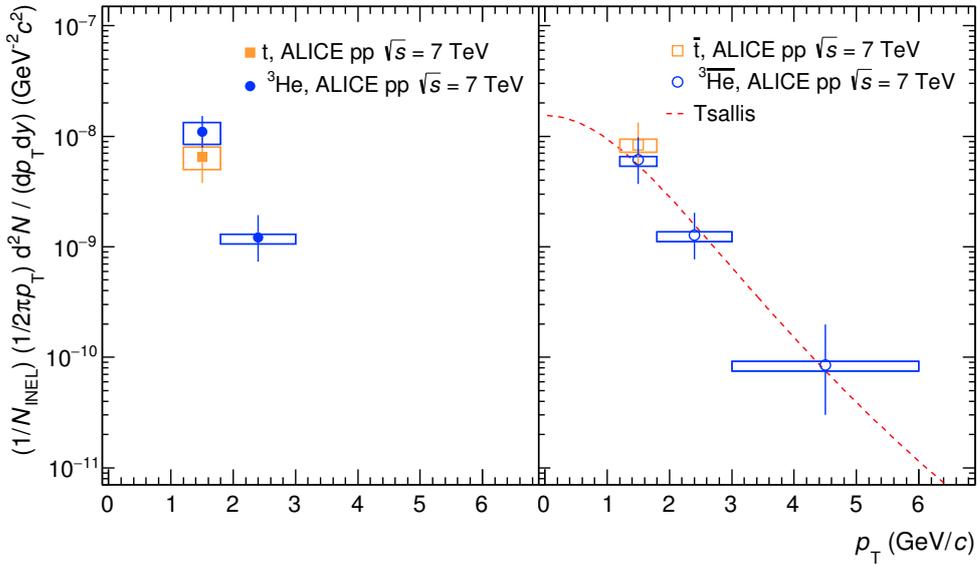}
\par\end{centering}

\caption{\label{fig:HeavNucSpectra}Invariant differential yields of tritons
and $^{3}$He nuclei (left panel) and their antinuclei (right panel)
in inelastic pp collisions at $\sqrt{s}$ = 7~TeV.
Error bars and boxes represent the statistical and systematic uncertainties,
respectively, and the dashed line the result of a fit with a Tsallis function
(see Sec.~\ref{sub:yields-mean-pt} for details).}
\end{figure}

A recorded luminosity of 4.2~nb$^{-1}$ allowed antitritons 
and $^{3}\overline{\text{He}}$ nuclei to be detected in pp collisions.
Since the total number of observed candidates is small, the uncertainties
were estimated as a central confidence interval (two-sided), using
a coverage probability of 68.27\% for a Poisson distribution. The
resulting invariant yields for both antinucleus species are compatible
in the $p_{\text{T}}$ range where measurements were possible 
(Fig.~\ref{fig:HeavNucSpectra}). Some $^{3}$He nuclei were also observed
in the highest $p_{\text{T}}$ bin, but, since the production rate is very
small, it was not feasible to evaluate the contamination due to secondary
$^{3}$He nuclei, and the bin was then excluded from this measurement.
In contrast, $^{3}\overline{\text{He}}$ nuclei are
not affected by this source of contamination, and the three measurements
are sufficient to determine the parameters of the Tsallis distribution
to extrapolate the yields (see Sec.~\ref{sub:yields-mean-pt}).

 % experimental results
\section{Discussion \label{sec:discussion}}

\subsection{Coalescence parameter \label{sub:coalescence-parameter}}

Many measurements of light nuclei have been successfully explained
as the result of the coalescence of protons and neutrons that are
nearby in phase-space \citep{Zupancic63,Gutbrod76}.
In this model, the production of a nucleus with mass number $A=N+Z$ is related to the
production of nucleons at equal momentum per nucleon by 
\begin{equation}
E_{A}\frac{\mathrm{d}^{3}N_{A}}{\mathrm{d}p_{A}^{3}}=B_{A}\left(E_{\mathrm{p}}\frac{\mathrm{d}^{3}N_{\mathrm{p}}}{\mathrm{d}p_{\mathrm{p}}^{3}}\right)^{Z}\left(E_{\mathrm{n}}\frac{\mathrm{d}^{3}N_{\mathrm{n}}}{\mathrm{d}p_{\mathrm{n}}^{3}}\right)^{N},\;\vec{p}_{\mathrm{p}}=\vec{p}_{\mathrm{n}}=\frac{\vec{p}_{A}}{A}\label{eq:BA}
\end{equation}
where\textbf{ }$B_{A}$ is called the coalescence parameter. This parameter
has been found to be constant at low transverse momentum
in light-particle collisions \citep{H1,ZEUS2007}. In contrast, in
AA collisions it has been reported that $B_{A}$ decreases with
increasing centrality of the collision, and for each centrality it increases
with $p_{\text{T}}$ \citep{E896,NA49,PHENIX2005,ALICE-dHe3-PbPb}.

\begin{figure}[H]
\noindent \begin{centering}
\includegraphics[scale=0.6]{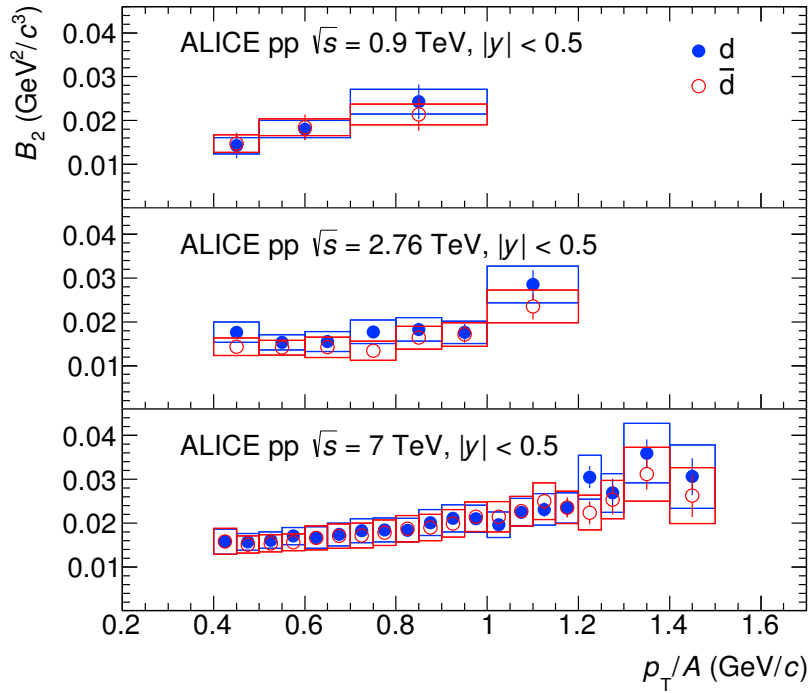}
\par\end{centering}

\caption{\label{fig:B2}Coalescence parameter ($B_{2}$) of deuterons (solid
circles) and antideuterons (hollow circles) as a function of $p_{\text{T}}$
per nucleon in inelastic pp collisions at $\sqrt{s}$ = 0.9, 2.76,
and 7~TeV. Statistical uncertainties are represented by error bars and
systematic uncertainties by boxes.}
\end{figure}

\begin{figure}[H]
\noindent \begin{centering}
\includegraphics[scale=0.6]{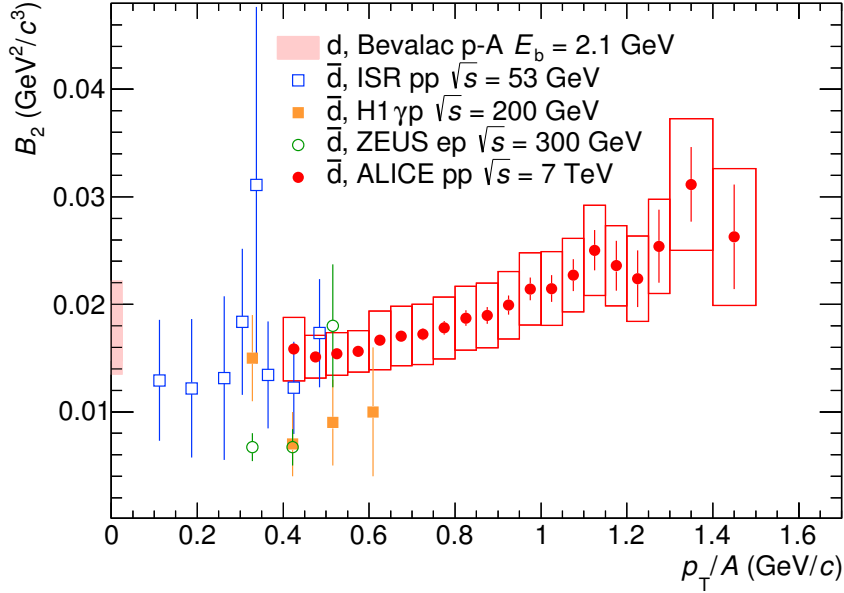}
\par\end{centering}

\caption{\label{fig:B2comp}Coalescence parameter ($B_{2}$) of antideuterons
in inelastic pp collisions at $\sqrt{s}$ = 7~TeV (circles) compared
with the values measured at lower energies in pp \citep{ISR,ISR-2}, $\upgamma$p 
\citep{H1}, ep \citep{ZEUS2007} (squares and hollow circles), and 
in p--Cu and p--Pb collisions \citep{Bevalac81}
(band at $p_{\text{T}}/A$~=~0~GeV/$c$).}
\end{figure}

Assuming equal distribution of nucleons in Eq.~(\ref{eq:BA}) and taking the proton and antiproton distributions
from Refs.~\citep{ALICE-piKp-900GeV,ALICE_pikp276,ALICE-piKp-7TeV},
the coalescence parameter ($B_{2}$) was computed, and it is shown in Fig.~\ref{fig:B2}.
The resulting values for deuterons and antideuterons are
compatible and do not show any significant dependence on the
center-of-mass energy within uncertainties.
These measurements extend the $p_{\text{T}}$
reach up to three times beyond previous measurements in pp collisions
extracted from the CERN Intersecting Storage Rings (ISR) \citep{ISR,ISR-2,Alper75}
(Fig.~\ref{fig:B2comp}).

To extract the $B_2$ from the CERN ISR, the antiproton distribution was taken
from \citep{Alper75} and the total cross section of $42.3\pm0.4$~mb from \citep{Ambrosio82}.
The distribution was also scaled by a factor of 0.69, estimated with an EPOS (LHC)
simulation \citep{Pierog2015,ALICE-PUBLIC-2017-010}, to take into account
the feed-down contribution. Figure \ref{fig:B2comp} also
includes the $B_2$ parameter of antideuterons from $\upgamma$p
collisions and deep inelastic scattering of electrons
at the Hadron-Electron Ring Accelerator (HERA) at DESY \citep{H1,ZEUS2007} 
and $B_2$ from p--Cu and p--Pb collisions at the LBNL
Bevalac \citep{Bevalac81}. Our measurement reveals a $p_{\text{T}}$
dependence in $B_{2}$ not seen in previous experiments, which is
significant given that the systematic uncertainties are correlated
bin by bin.

\begin{figure}[H]
\noindent \begin{centering}
\includegraphics[scale=0.6]{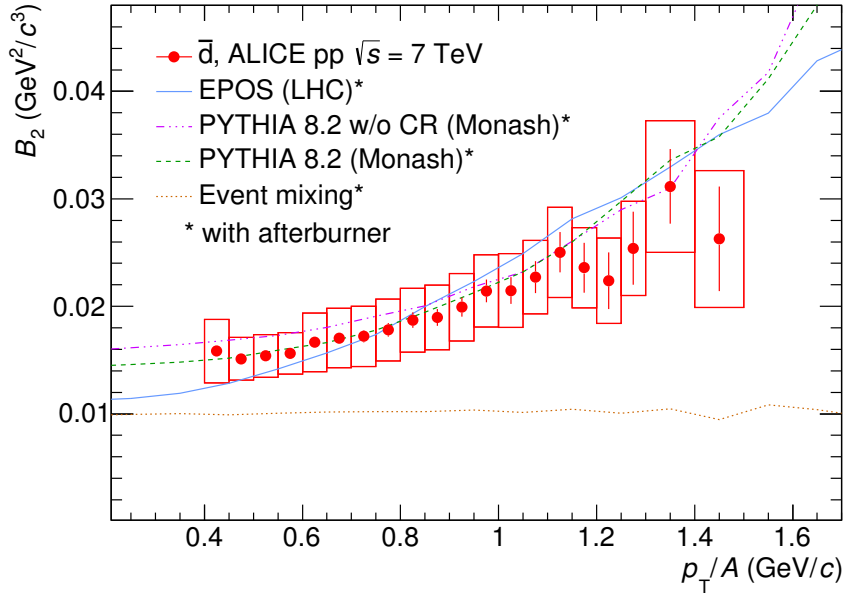}
\par\end{centering}

\caption{\label{fig:B2MC} Coalescence parameter ($B_{2}$) of antideuterons 
in inelastic pp collisions at $\sqrt{s}$ = 7~TeV (circles) compared with EPOS (LHC), 
PYTHIA 8.2 (Monash tune) with and without color reconnection (CR), and an 
event mixing procedure with the afterburner (lines).}
\end{figure}

This $p_{\text{T}}$ dependence can be reproduced with
QCD-inspired event generators, such as PYTHIA 8.2 (Monash tune) \citep{Sjostrand2008}
and EPOS (LHC), when adding a coalescence-based afterburner \citep{ALICE-PUBLIC-2017-010} that
takes into account the momentum correlations between nucleons (Fig.~\ref{fig:B2MC}).
The afterburner looks for clusters of nucleons among the final particles
produced by the event generators and boosts them to their center-of-mass frame.
If the momentum of each individual nucleon is less than a certain value,
a nucleus is generated. 
With the afterburner, a constant $B_{2}$ is recovered when selecting protons 
from one event and neutrons from the next event (event mixing), in agreement with the expectation
of an uncorrelated distribution of nucleons (Fig.~\ref{fig:B2MC}).
The $p_{\text{T}}$ dependence in $B_{2}$ is still present in the results from an alternate
PYTHIA 8.2 (Monash tune) simulation with color reconnection turned off (Fig.~\ref{fig:B2MC}).
Furthermore, a radial flow effect in $B_{2}$ at these low average
charged multiplicities is also discarded by the EPOS (LHC)
simulation with the afterburner, since this contribution only arises
in high multiplicity events, starting from $\text{d}N_{ch}/\text{d}\eta>15$ \citep{Pierog2015}.
Thus, this $p_{\text{T}}$ dependence can be explained as
a purely hard scattering effect, in contrast to AA collisions, where it is usually attributed
to collective flow.

\begin{figure}[H]
\noindent \begin{centering}
\includegraphics[scale=0.7]{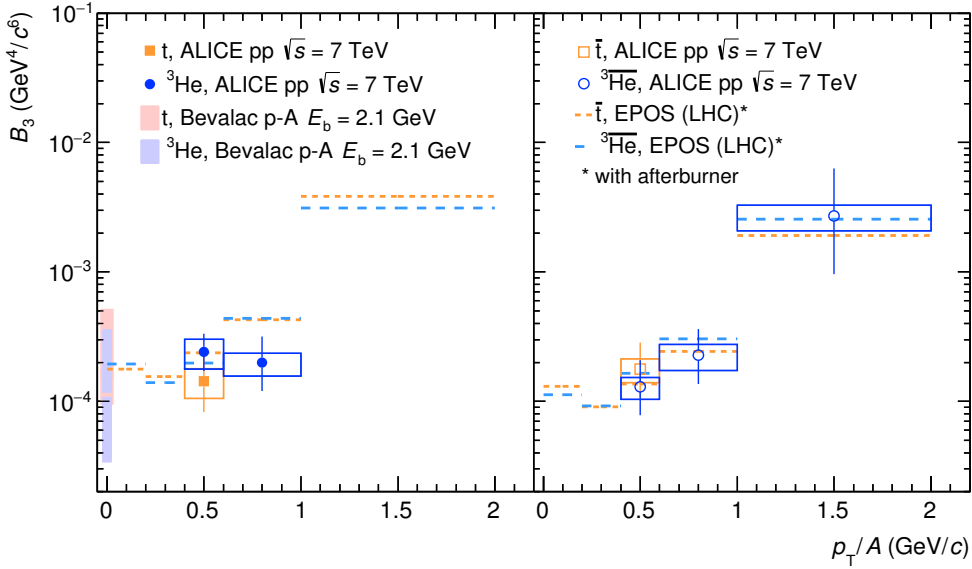}
\par\end{centering}

\caption{\label{fig:B3}Coalescence parameter ($B_{3}$) of tritons and $^{3}$He
nuclei (left panel) and their antinuclei (right panel) in inelastic
pp collisions at $\sqrt{s}$~=~7~TeV. The Bevalac measurements in p--C, p--Cu,
and p--Pb collisions \citep{Bevalac81} are not given as a function of $p_{\text{T}}$
and are shown as vertical bands at $p_{\text{T}}/A$~=~0~GeV/$c$ for comparison.
Error bars and boxes represent the statistical and systematic uncertainties,
respectively, and dashed lines the values obtained with EPOS (LHC) with
the afterburner.}
\end{figure}

As in the case of antideuterons, the coalescence parameter ($B_{3}$)
of $^{3}\overline{\text{He}}$ nuclei also exhibits a $p_{\text{T}}$
dependence (Fig.~\ref{fig:B3} right), and can be reproduced with QCD-inspired
event generators with a coalescence-based afterburner \citep{ALICE-PUBLIC-2017-010}.
Moreover, low $p_{\text{T}}$ values of $B_{3}$ are compatible with
those obtained in p--C, p--Cu, and p--Pb collisions at Bevalac \citep{Bevalac81}. 

\subsection{Integrated yields and deuteron-to-proton ratio \label{sub:yields-mean-pt}}

Unlike coalescence models, statistical hadronization models only provide
predictions for integrated yields. In this case, the integrated yields of light
nuclei and the deuteron-to-proton ratio can add additional constraints to these
models and could therefore serve as a test for thermal-statistical behavior 
in small systems at LHC energies.

To find the integrated yields, the measurements were extrapolated to the
unmeasured $p_{\text{T}}$ region with a statistical distribution that provides
an exponential behavior at low $p_{\text{T}}$ and a power law behavior at high
$p_{\text{T}}$ (Figs.~\ref{fig:DbarDSpectra} and \ref{fig:HeavNucSpectra}):
\begin{equation}
E\frac{\text{d}^{3}N}{\text{d}p^{3}}=gV\frac{m_{\text{T}}}{(2\pi)^{3}}\left(1+(q-1)\frac{m_{\text{T}}}{T}\right)^{\frac{q}{1-q}}\label{eq:diff-inv-yield-Tallis},
\end{equation}
where $m_{\text{T}}=\sqrt{p_{\text{T}}^{2}+m^{2}}$ is the transverse mass,
and $gV$, $T$, and $q$ are free parameters. This distribution can be derived
from the Tsallis entropy \citep{Tsallis88,Cleymans2012} and gives good
description of the data in pp collisions \citep{Cleymans2012}. It was preferred
over the Levy-Tsallis used in previous work \citep{ALICE-dHe3-PbPb} as
it provides a more stable description of the measurements with a limited
data set, as in the case of antideuterons for the center-of-mass energy 0.9~TeV
or the $^{3}\overline{\text{He}}$ nuclei.

\begin{figure}[H]
\noindent \begin{centering}
\includegraphics[scale=0.6]{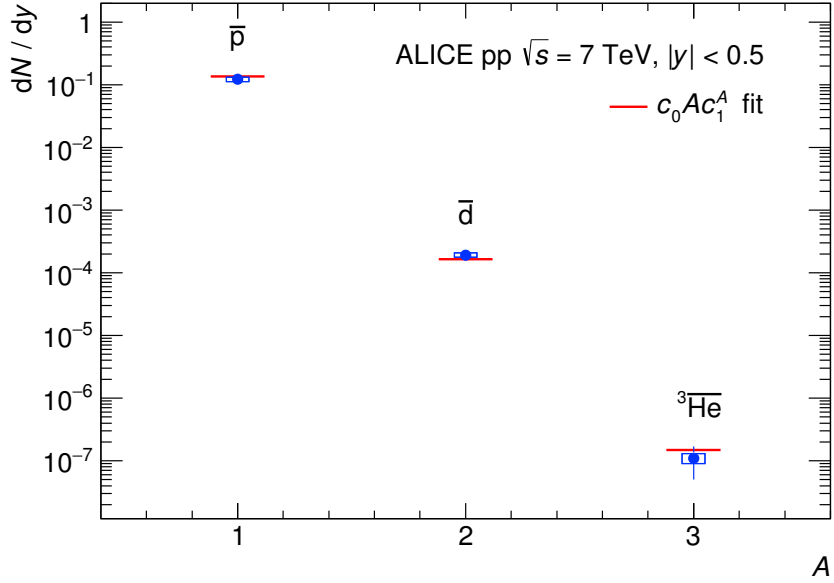}
\par\end{centering}

\caption{\label{fig:dNdy} Integrated yields ($\text{d}N/\text{d}y$) of antiprotons,
antideuterons and $^{3}\overline{\text{He}}$ nuclei
as a function of the number of antinucleons in inelastic pp collisions
at $\sqrt{s}$ = 7~TeV. The horizontal lines represent a fit with the function
$c_{0}Ac_{1}^{A}$ based on Eq.~(\ref{eq:BA}).}
\end{figure}

The systematic uncertainties of the integrated yields ($\text{d}N/\text{d}y$) and
mean transverse momenta ($\left\langle p_{\text{T}}\right\rangle $) were evaluated
by shifting the data points up and then down by their uncertainties
(i.e., assuming full correlation between $p_{\text{T}}$ bins). Additionally, 
the data points were shifted coherently, in a $p_{\text{T}}$-dependent way,
within their uncertainties to create maximally hard and maximally soft $p_{\text{T}}$ distributions.
The values of $\text{d}N/\text{d}y$ and $\left\langle p_{\text{T}}\right\rangle $ 
were reevaluated and the largest difference was taken as the systematic uncertainty.
Table~\ref{tab:dNdy} summarizes the resulting values for the different center-of-mass
energies along with the extrapolation fraction due to the unmeasured $p_{\text{T}}$ regions.
The uncertainty on the extrapolation was estimated by using additional distributions
including the Levy-Tsallis \citep{Abelev2007,Adare2011} and 
Boltzmann distributions.
The change of the default fit function with respect to 
\citep{ALICE-dHe3-PbPb} leads to slightly
different values for the obtained $\text{d}N/\text{d}y$ and 
$\left\langle p_{\text{T}}\right\rangle $ which are consistent within
the respective systematic uncertainties. Figure \ref{fig:dNdy} shows an exponential
decrease of the $\text{d}N/\text{d}y$ as a function of the mass number.
The reduction of the yield for each additional nucleon is about 1000.

\begin{table}[H]
\begin{centering}
{\footnotesize{}}%
\begin{tabular}{ccccc}
\hline 
 & $\sqrt{s}$ (TeV) & $\text{d}N/\text{d}y$ & $\left\langle p_{\text{T}}\right\rangle $ (GeV/$c$) & Extr.\tabularnewline
\hline 
 & 0.9 & ( 1.12 $\pm$ 0.09 $\pm$ 0.09 ) $\times10^{-4}$ & 1.01 $\pm$ 0.05 $\pm$ 0.05  & 50 $\pm$ 3\%\tabularnewline
d & 2.76 & ( 1.53 $\pm$ 0.05 $\pm$ 0.13 ) $\times10^{-4}$ & 1.04 $\pm$ 0.02 $\pm$ 0.04  & 45 $\pm$ 8\%\tabularnewline
 & 7 & ( 2.02 $\pm$ 0.02 $\pm$ 0.17 ) $\times10^{-4}$ & 1.11 $\pm$ 0.01 $\pm$ 0.04  & 41 $\pm$ 5\%\tabularnewline
\hline 
 & 0.9 & ( 1.11 $\pm$ 0.10 $\pm$ 0.09 ) $\times10^{-4}$ & 0.99 $\pm$ 0.07 $\pm$ 0.05  & 52 $\pm$ 7\%\tabularnewline
$\overline{\text{d}}$ & 2.76 & ( 1.37 $\pm$ 0.04 $\pm$ 0.12 ) $\times10^{-4}$ & 1.04 $\pm$ 0.02 $\pm$ 0.03 & 46 $\pm$ 7\%\tabularnewline
 & 7 & ( 1.92 $\pm$ 0.02 $\pm$ 0.15 ) $\times10^{-4}$ & 1.08 $\pm$ 0.01 $\pm$ 0.04  & 42 $\pm$ 5\%\tabularnewline
\hline 
$^{3}\overline{\text{He}}$ & 7 & ( 1.1 $\pm$ 0.6 $\pm$ 0.2 ) $\times10^{-7}$ & 1.6 $\pm$ 0.4 $\pm$ 0.04  & 43 $\pm$ 14\%\tabularnewline
\hline 
\end{tabular}
\par\end{centering}{\footnotesize \par}

\caption{\label{tab:dNdy} Integrated yields ($\text{d}N/\text{d}y$)
and mean transverse momenta ($\left\langle p_{\text{T}}\right\rangle $)
for deuterons, antideuterons, and $^{3}\overline{\text{He}}$ nuclei
along with the extrapolated fraction (Extr.) due to the unmeasured
$p_{\text{T}}$ regions. The first uncertainty is the statistical
uncertainty and the second one the systematic uncertainty.}
 
\end{table}

\begin{figure}[H]
\noindent \begin{centering}
\includegraphics[scale=0.6]{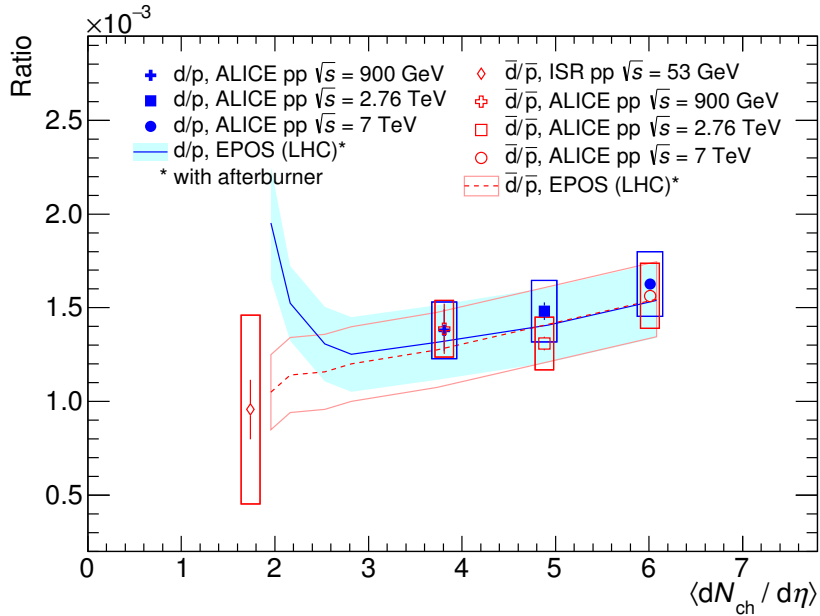}
\par\end{centering}

\caption{\label{fig:DbarPbarRatio} Integrated deuteron-to-proton (d/p) 
and antideuteron-to-antiproton ($\overline{\text{d}}/\overline{\text{p}}$)
ratios in inelastic pp collisions as a function of the average
charged particle multiplicity for different center-of-mass energies.
The average d/p ratio in AA collisions lies two times above the highest value 
in pp collisions (not shown). Dashed and solid lines represent the expected 
values from EPOS (LHC) with afterburner and the bands their uncertainties.
The CERN ISR value is corrected by the contribution of feed-down
antiprotons estimated with an EPOS (LHC) simulation.}
\end{figure}

The integrated d/p and $\overline{\text{d}}/\overline{\text{p}}$
ratios were calculated from the integrated yields in Table \ref{tab:dNdy}
and Refs.~\citep{ALICE_pikp276,ALICE-piKp-7TeV}, and are shown in
Fig.~\ref{fig:DbarPbarRatio} as a function of the average charged
particle multiplicity at mid-rapidity \citep{Thome77,ALICE-pp7mult}.
The $\text{d}N/\text{d}y$ values for pp collisions at the CERN ISR
were computed following the same procedure described above and
using the inclusive $\overline{\text{p}}$ distribution from \citep{Alper75}
and the $\overline{\text{d}}$ distribution from Refs.~\citep{ISR,ISR-2}.
The resulting $\overline{\text{d}}/\overline{\text{p}}$ ratio was
divided by 0.69 to account for the contributions of feed-down antiprotons,
based on an EPOS (LHC) simulation
\citep{ALICE-PUBLIC-2017-010}. Figure \ref{fig:DbarPbarRatio} suggests
an increasing trend in this ratio with the average charged particle multiplicity
in pp collisions, which is also supported by an EPOS (LHC)
simulation with the afterburner, although at ISR energies the d/p ratio
is strongly influenced by the baryon number transport at mid-rapidity leading
to a higher value than at LHC energies according to the model expectations.

When describing particle ratios such as the d/p ratio, the only free parameter
of grand-canonical statistical hadronization models at LHC energies is the
chemical freeze-out temperature.
In the past, several attempts were made to extend their successful
description of AA collisions to smaller collision systems such as pp. In particular, 
the canonical formulation describes the production of light flavor hadrons, 
including those with strangeness content \citep{Andronic2011}.
While the p/$\uppi$ ratio is found to be comparable in pp, p--Pb, and Pb--Pb collisions
\citep{ALICE-piKp-7TeV,ALICE-piKp-pPb-2014}, indicating a comparable chemical
freeze-out temperature among different systems, the d/p ratio in pp collisions
at LHC energies is found to be two times lower than the average value in
AA collisions. Since the strangeness-canonical 
formulation only influences the abundance of strange particles with respect to
nonstrange particles, it cannot explain the observed results presented here.

 % discussion
\section{Summary and conclusions \label{sec:summary-conclusions}}

The invariant differential yields of deuterons and antideuterons
in pp collisions at $\sqrt{s}$ = 0.9, 2.76, and 7~TeV and the 
yields of tritons, $^{3}$He
nuclei, and their antinuclei at $\sqrt{s}$ = 7~TeV have been measured
in the rapidity range $|y|<0.5$. The measurements cover the $p_{\text{T}}$
ranges $0.8<p_{\text{T}}<3$~GeV/$c$ for (anti)deuterons, $1.2<p_{\text{T}}<1.8$~GeV/$c$
for (anti)tritons, $1.2<p_{\text{T}}<3$~GeV/$c$ for $^{3}$He nuclei,
and $1.2<p_{\text{T}}<6$~GeV/$c$ for $^{3}$He antinuclei.
This extends previous measurements by one order of
magnitude in incident energies, and by a factor of 3 in $p_{\text{T}}$
reach, and it includes the first ever measurements of antitritons and
$^{3}\overline{\text{He}}$ nuclei in pp collisions.

The present measurements show no significant dependence of the coalescence 
parameter $B_{2}$ on the center-of-mass energy from CERN ISR energies
(53~GeV) to the highest LHC energy reported in this paper (7~TeV).
Moreover, the values of both $B_{2}$ and $B_{3}$ are found to be compatible
at low $p_{\text{T}}$ with those obtained in pA collisions at Bevalac energies.

A previously unobserved $p_{\text{T}}$ dependence in pp collisions
of the coalescence parameters $B_{2}$ and $B_{3}$ is also reported.
The data are well described by QCD-inspired event generators when
a coalescence-based afterburner is added to take into account the
momentum correlations between nucleons. According to
PYTHIA 8.2 (Monash tune) and EPOS (LHC) with the afterburner,
this dependence can be explained purely as a hard scattering effect.

In combination with CERN ISR measurements, our results suggest an
increasing trend in the $\overline{\text{d}}/\overline{\text{p}}$
ratio with charged particle multiplicity. While the values reported
in central AA collisions are in agreement with a thermal model description
of particle yields, the highest d/p ratio reported in this paper is about half
the thermal model value;
therefore, a thermal-statistical description is disfavored in pp collisions
at these low average charged particle multiplicities.
Our measurements are expected to contribute to the understanding of
the background from pp collisions for the observation of antideuterons
and $^{3}\overline{\text{He}}$ nuclei in cosmic ray experiments and
to the estimation of the production rates of the next stable antinuclei
in pp collisions.

 % summary and conclusions
%
%

%%%%% acknowledgements
\newenvironment{acknowledgement}{\relax}{\relax}
\begin{acknowledgement}
\section*{Acknowledgements}
% Version: 2017-09-22

The ALICE Collaboration would like to thank all its engineers and technicians for their invaluable contributions to the construction of the experiment and the CERN accelerator teams for the outstanding performance of the LHC complex.
The ALICE Collaboration gratefully acknowledges the resources and support provided by all Grid centres and the Worldwide LHC Computing Grid (WLCG) collaboration.
The ALICE Collaboration acknowledges the following funding agencies for their support in building and running the ALICE detector:
A. I. Alikhanyan National Science Laboratory (Yerevan Physics Institute) Foundation (ANSL), State Committee of Science and World Federation of Scientists (WFS), Armenia;
Austrian Academy of Sciences and Nationalstiftung f\"{u}r Forschung, Technologie und Entwicklung, Austria;
Ministry of Communications and High Technologies, National Nuclear Research Center, Azerbaijan;
Conselho Nacional de Desenvolvimento Cient\'{\i}fico e Tecnol\'{o}gico (CNPq), Universidade Federal do Rio Grande do Sul (UFRGS), Financiadora de Estudos e Projetos (Finep) and Funda\c{c}\~{a}o de Amparo \`{a} Pesquisa do Estado de S\~{a}o Paulo (FAPESP), Brazil;
Ministry of Science \& Technology of China (MSTC), National Natural Science Foundation of China (NSFC) and Ministry of Education of China (MOEC) , China;
Ministry of Science, Education and Sport and Croatian Science Foundation, Croatia;
Ministry of Education, Youth and Sports of the Czech Republic, Czech Republic;
The Danish Council for Independent Research | Natural Sciences, the Carlsberg Foundation and Danish National Research Foundation (DNRF), Denmark;
Helsinki Institute of Physics (HIP), Finland;
Commissariat \`{a} l'Energie Atomique (CEA) and Institut National de Physique Nucl\'{e}aire et de Physique des Particules (IN2P3) and Centre National de la Recherche Scientifique (CNRS), France;
Bundesministerium f\"{u}r Bildung, Wissenschaft, Forschung und Technologie (BMBF) and GSI Helmholtzzentrum f\"{u}r Schwerionenforschung GmbH, Germany;
General Secretariat for Research and Technology, Ministry of Education, Research and Religions, Greece;
National Research, Development and Innovation Office, Hungary;
Department of Atomic Energy Government of India (DAE), Department of Science and Technology, Government of India (DST), University Grants Commission, Government of India (UGC) and Council of Scientific and Industrial Research (CSIR), India;
Indonesian Institute of Science, Indonesia;
Centro Fermi - Museo Storico della Fisica e Centro Studi e Ricerche Enrico Fermi and Istituto Nazionale di Fisica Nucleare (INFN), Italy;
Institute for Innovative Science and Technology , Nagasaki Institute of Applied Science (IIST), Japan Society for the Promotion of Science (JSPS) KAKENHI and Japanese Ministry of Education, Culture, Sports, Science and Technology (MEXT), Japan;
Consejo Nacional de Ciencia (CONACYT) y Tecnolog\'{i}a, through Fondo de Cooperaci\'{o}n Internacional en Ciencia y Tecnolog\'{i}a (FONCICYT) and Direcci\'{o}n General de Asuntos del Personal Academico (DGAPA), Mexico;
Nederlandse Organisatie voor Wetenschappelijk Onderzoek (NWO), Netherlands;
The Research Council of Norway, Norway;
Commission on Science and Technology for Sustainable Development in the South (COMSATS), Pakistan;
Pontificia Universidad Cat\'{o}lica del Per\'{u}, Peru;
Ministry of Science and Higher Education and National Science Centre, Poland;
Korea Institute of Science and Technology Information and National Research Foundation of Korea (NRF), Republic of Korea;
Ministry of Education and Scientific Research, Institute of Atomic Physics and Romanian National Agency for Science, Technology and Innovation, Romania;
Joint Institute for Nuclear Research (JINR), Ministry of Education and Science of the Russian Federation and National Research Centre Kurchatov Institute, Russia;
Ministry of Education, Science, Research and Sport of the Slovak Republic, Slovakia;
National Research Foundation of South Africa, South Africa;
Centro de Aplicaciones Tecnol\'{o}gicas y Desarrollo Nuclear (CEADEN), Cubaenerg\'{\i}a, Cuba, Ministerio de Ciencia e Innovacion and Centro de Investigaciones Energ\'{e}ticas, Medioambientales y Tecnol\'{o}gicas (CIEMAT), Spain;
Swedish Research Council (VR) and Knut \& Alice Wallenberg Foundation (KAW), Sweden;
European Organization for Nuclear Research, Switzerland;
National Science and Technology Development Agency (NSDTA), Suranaree University of Technology (SUT) and Office of the Higher Education Commission under NRU project of Thailand, Thailand;
Turkish Atomic Energy Agency (TAEK), Turkey;
National Academy of  Sciences of Ukraine, Ukraine;
Science and Technology Facilities Council (STFC), United Kingdom;
National Science Foundation of the United States of America (NSF) and United States Department of Energy, Office of Nuclear Physics (DOE NP), United States of America.    %%%%%%% done by webmaster team
\end{acknowledgement}

%%%%%%%% Bibliography (In case of using bibtex generate the bbl requested by arXiv)
%\bibliographystyle{style}   % Put here the style file name for the paper, i.e.apsrev4-1, utphys
%\bibliography{biblio}
%\input {bibliography.tex}  

\bibliographystyle{utphys}
\bibliography{references}

%%%%%%%%% appendix with author list
\newpage
\appendix
\section{The ALICE Collaboration}
\label{app:collab}
% Collaboration: CERN-LHC-ALICE
% Generation Date is 2017-Aug-18

% How to use:
%%%%%%%%% appendix with author list
%\appendix
%\section{The ALICE Collaboration}
%\label{app:collab}
%\input{Alice_Authorslist_XXXX-Axx-XX.tex}
\begingroup
\small
\begin{flushleft}
S.~Acharya\Irefn{org137}\And 
J.~Adam\Irefn{org96}\And 
D.~Adamov\'{a}\Irefn{org93}\And 
J.~Adolfsson\Irefn{org32}\And 
M.M.~Aggarwal\Irefn{org98}\And 
G.~Aglieri Rinella\Irefn{org33}\And 
M.~Agnello\Irefn{org29}\And 
N.~Agrawal\Irefn{org46}\And 
Z.~Ahammed\Irefn{org137}\And 
N.~Ahmad\Irefn{org15}
% ...
\And
\renewcommand\labelenumi{\textsuperscript{\theenumi}~}

\section*{Affiliation notes}
\renewcommand\theenumi{\roman{enumi}}
\begin{Authlist}
\item \Adef{org*}Deceased
\item \Adef{orgI}Dipartimento DET del Politecnico di Torino, Turin, Italy
\item \Adef{orgII}Georgia State University, Atlanta, Georgia, United States
\item \Adef{orgIII}M.V. Lomonosov Moscow State University, D.V. Skobeltsyn Institute of Nuclear, Physics, Moscow, Russia
\item \Adef{orgIV}Department of Applied Physics, Aligarh Muslim University, Aligarh, India
\item \Adef{orgV}Institute of Theoretical Physics, University of Wroclaw, Poland
\end{Authlist}

\section*{Collaboration Institutes}
\renewcommand\theenumi{\arabic{enumi}~}
\begin{Authlist}
\item \Idef{org1}A.I. Alikhanyan National Science Laboratory (Yerevan Physics Institute) Foundation, Yerevan, Armenia
\item \Idef{org2}Benem\'{e}rita Universidad Aut\'{o}noma de Puebla, Puebla, Mexico
\item \Idef{org3}Bogolyubov Institute for Theoretical Physics, Kiev, Ukraine
\item \Idef{org4}Bose Institute, Department of Physics  and Centre for Astroparticle Physics and Space Science (CAPSS), Kolkata, India
\item \Idef{org5}Budker Institute for Nuclear Physics, Novosibirsk, Russia
\item \Idef{org6}California Polytechnic State University, San Luis Obispo, California, United States
\item \Idef{org7}Central China Normal University, Wuhan, China
\item \Idef{org8}Centre de Calcul de l'IN2P3, Villeurbanne, Lyon, France
\item \Idef{org9}Centro de Aplicaciones Tecnol\'{o}gicas y Desarrollo Nuclear (CEADEN), Havana, Cuba
\item \Idef{org10}Centro de Investigaci\'{o}n y de Estudios Avanzados (CINVESTAV), Mexico City and M\'{e}rida, Mexico
\item \Idef{org11}Centro Fermi - Museo Storico della Fisica e Centro Studi e Ricerche ``Enrico Fermi', Rome, Italy
\item \Idef{org12}Chicago State University, Chicago, Illinois, United States
\item \Idef{org13}China Institute of Atomic Energy, Beijing, China
\item \Idef{org14}COMSATS Institute of Information Technology (CIIT), Islamabad, Pakistan
\item \Idef{org15}Department of Physics, Aligarh Muslim University, Aligarh, India
\item \Idef{org16}Department of Physics, Ohio State University, Columbus, Ohio, United States
\item \Idef{org17}Department of Physics, Pusan National University, Pusan, Republic of Korea
\item \Idef{org18}Department of Physics, Sejong University, Seoul, Republic of Korea
\item \Idef{org19}Department of Physics, University of Oslo, Oslo, Norway
\item \Idef{org20}Department of Physics and Technology, University of Bergen, Bergen, Norway
\item \Idef{org21}Dipartimento di Fisica dell'Universit\`{a} 'La Sapienza' and Sezione INFN, Rome, Italy
\item \Idef{org22}Dipartimento di Fisica dell'Universit\`{a} and Sezione INFN, Cagliari, Italy
\item \Idef{org23}Dipartimento di Fisica dell'Universit\`{a} and Sezione INFN, Trieste, Italy
\item \Idef{org24}Dipartimento di Fisica dell'Universit\`{a} and Sezione INFN, Turin, Italy
\item \Idef{org25}Dipartimento di Fisica e Astronomia dell'Universit\`{a} and Sezione INFN, Bologna, Italy
\item \Idef{org26}Dipartimento di Fisica e Astronomia dell'Universit\`{a} and Sezione INFN, Catania, Italy
\item \Idef{org27}Dipartimento di Fisica e Astronomia dell'Universit\`{a} and Sezione INFN, Padova, Italy
\item \Idef{org28}Dipartimento di Fisica `E.R.~Caianiello' dell'Universit\`{a} and Gruppo Collegato INFN, Salerno, Italy
\item \Idef{org29}Dipartimento DISAT del Politecnico and Sezione INFN, Turin, Italy
\item \Idef{org30}Dipartimento di Scienze e Innovazione Tecnologica dell'Universit\`{a} del Piemonte Orientale and INFN Sezione di Torino, Alessandria, Italy
\item \Idef{org31}Dipartimento Interateneo di Fisica `M.~Merlin' and Sezione INFN, Bari, Italy
\item \Idef{org32}Division of Experimental High Energy Physics, University of Lund, Lund, Sweden
\item \Idef{org33}European Organization for Nuclear Research (CERN), Geneva, Switzerland
\item \Idef{org34}Excellence Cluster Universe, Technische Universit\"{a}t M\"{u}nchen, Munich, Germany
\item \Idef{org35}Faculty of Engineering, Bergen University College, Bergen, Norway
\item \Idef{org36}Faculty of Mathematics, Physics and Informatics, Comenius University, Bratislava, Slovakia
\item \Idef{org37}Faculty of Nuclear Sciences and Physical Engineering, Czech Technical University in Prague, Prague, Czech Republic
\item \Idef{org38}Faculty of Science, P.J.~\v{S}af\'{a}rik University, Ko\v{s}ice, Slovakia
\item \Idef{org39}Faculty of Technology, Buskerud and Vestfold University College, Tonsberg, Norway
\item \Idef{org40}Frankfurt Institute for Advanced Studies, Johann Wolfgang Goethe-Universit\"{a}t Frankfurt, Frankfurt, Germany
\item \Idef{org41}Gangneung-Wonju National University, Gangneung, Republic of Korea
\item \Idef{org42}Gauhati University, Department of Physics, Guwahati, India
\item \Idef{org43}Helmholtz-Institut f\"{u}r Strahlen- und Kernphysik, Rheinische Friedrich-Wilhelms-Universit\"{a}t Bonn, Bonn, Germany
\item \Idef{org44}Helsinki Institute of Physics (HIP), Helsinki, Finland
\item \Idef{org45}Hiroshima University, Hiroshima, Japan
\item \Idef{org46}Indian Institute of Technology Bombay (IIT), Mumbai, India
\item \Idef{org47}Indian Institute of Technology Indore, Indore, India
\item \Idef{org48}Indonesian Institute of Sciences, Jakarta, Indonesia
\item \Idef{org49}INFN, Laboratori Nazionali di Frascati, Frascati, Italy
\item \Idef{org50}INFN, Laboratori Nazionali di Legnaro, Legnaro, Italy
\item \Idef{org51}INFN, Sezione di Bari, Bari, Italy
\item \Idef{org52}INFN, Sezione di Bologna, Bologna, Italy
\item \Idef{org53}INFN, Sezione di Cagliari, Cagliari, Italy
\item \Idef{org54}INFN, Sezione di Catania, Catania, Italy
\item \Idef{org55}INFN, Sezione di Padova, Padova, Italy
\item \Idef{org56}INFN, Sezione di Roma, Rome, Italy
\item \Idef{org57}INFN, Sezione di Torino, Turin, Italy
\item \Idef{org58}INFN, Sezione di Trieste, Trieste, Italy
\item \Idef{org59}Inha University, Incheon, Republic of Korea
\item \Idef{org60}Institut de Physique Nucl\'eaire d'Orsay (IPNO), Universit\'e Paris-Sud, CNRS-IN2P3, Orsay, France
\item \Idef{org61}Institute for Nuclear Research, Academy of Sciences, Moscow, Russia
\item \Idef{org62}Institute for Subatomic Physics of Utrecht University, Utrecht, Netherlands
\item \Idef{org63}Institute for Theoretical and Experimental Physics, Moscow, Russia
\item \Idef{org64}Institute of Experimental Physics, Slovak Academy of Sciences, Ko\v{s}ice, Slovakia
\item \Idef{org65}Institute of Physics, Academy of Sciences of the Czech Republic, Prague, Czech Republic
\item \Idef{org66}Institute of Physics, Bhubaneswar, India
\item \Idef{org67}Institute of Space Science (ISS), Bucharest, Romania
\item \Idef{org68}Institut f\"{u}r Informatik, Johann Wolfgang Goethe-Universit\"{a}t Frankfurt, Frankfurt, Germany
\item \Idef{org69}Institut f\"{u}r Kernphysik, Johann Wolfgang Goethe-Universit\"{a}t Frankfurt, Frankfurt, Germany
\item \Idef{org70}Institut f\"{u}r Kernphysik, Westf\"{a}lische Wilhelms-Universit\"{a}t M\"{u}nster, M\"{u}nster, Germany
\item \Idef{org71}Instituto de Ciencias Nucleares, Universidad Nacional Aut\'{o}noma de M\'{e}xico, Mexico City, Mexico
\item \Idef{org72}Instituto de F\'{i}sica, Universidade Federal do Rio Grande do Sul (UFRGS), Porto Alegre, Brazil
\item \Idef{org73}Instituto de F\'{\i}sica, Universidad Nacional Aut\'{o}noma de M\'{e}xico, Mexico City, Mexico
\item \Idef{org74}IRFU, CEA, Universit\'{e} Paris-Saclay, Saclay, France
\item \Idef{org75}iThemba LABS, National Research Foundation, Somerset West, South Africa
\item \Idef{org76}Joint Institute for Nuclear Research (JINR), Dubna, Russia
\item \Idef{org77}Konkuk University, Seoul, Republic of Korea
\item \Idef{org78}Korea Institute of Science and Technology Information, Daejeon, Republic of Korea
\item \Idef{org79}KTO Karatay University, Konya, Turkey
\item \Idef{org80}Laboratoire de Physique Subatomique et de Cosmologie, Universit\'{e} Grenoble-Alpes, CNRS-IN2P3, Grenoble, France
\item \Idef{org81}Lawrence Berkeley National Laboratory, Berkeley, California, United States
\item \Idef{org82}Moscow Engineering Physics Institute, Moscow, Russia
\item \Idef{org83}Nagasaki Institute of Applied Science, Nagasaki, Japan
\item \Idef{org84}National and Kapodistrian University of Athens, Physics Department, Athens, Greece
\item \Idef{org85}National Centre for Nuclear Studies, Warsaw, Poland
\item \Idef{org86}National Institute for Physics and Nuclear Engineering, Bucharest, Romania
\item \Idef{org87}National Institute of Science Education and Research, HBNI, Jatni, India
\item \Idef{org88}National Nuclear Research Center, Baku, Azerbaijan
\item \Idef{org89}National Research Centre Kurchatov Institute, Moscow, Russia
\item \Idef{org90}Niels Bohr Institute, University of Copenhagen, Copenhagen, Denmark
\item \Idef{org91}Nikhef, Nationaal instituut voor subatomaire fysica, Amsterdam, Netherlands
\item \Idef{org92}Nuclear Physics Group, STFC Daresbury Laboratory, Daresbury, United Kingdom
\item \Idef{org93}Nuclear Physics Institute, Academy of Sciences of the Czech Republic, \v{R}e\v{z} u Prahy, Czech Republic
\item \Idef{org94}Oak Ridge National Laboratory, Oak Ridge, Tennessee, United States
\item \Idef{org95}Petersburg Nuclear Physics Institute, Gatchina, Russia
\item \Idef{org96}Physics Department, Creighton University, Omaha, Nebraska, United States
\item \Idef{org97}Physics department, Faculty of science, University of Zagreb, Zagreb, Croatia
\item \Idef{org98}Physics Department, Panjab University, Chandigarh, India
\item \Idef{org99}Physics Department, University of Cape Town, Cape Town, South Africa
\item \Idef{org100}Physics Department, University of Jammu, Jammu, India
\item \Idef{org101}Physics Department, University of Rajasthan, Jaipur, India
\item \Idef{org102}Physikalisches Institut, Eberhard Karls Universit\"{a}t T\"{u}bingen, T\"{u}bingen, Germany
\item \Idef{org103}Physikalisches Institut, Ruprecht-Karls-Universit\"{a}t Heidelberg, Heidelberg, Germany
\item \Idef{org104}Physik Department, Technische Universit\"{a}t M\"{u}nchen, Munich, Germany
\item \Idef{org105}Purdue University, West Lafayette, Indiana, United States
\item \Idef{org106}Research Division and ExtreMe Matter Institute EMMI, GSI Helmholtzzentrum f\"ur Schwerionenforschung GmbH, Darmstadt, Germany
\item \Idef{org107}Rudjer Bo\v{s}kovi\'{c} Institute, Zagreb, Croatia
\item \Idef{org108}Russian Federal Nuclear Center (VNIIEF), Sarov, Russia
\item \Idef{org109}Saha Institute of Nuclear Physics, Kolkata, India
\item \Idef{org110}School of Physics and Astronomy, University of Birmingham, Birmingham, United Kingdom
\item \Idef{org111}Secci\'{o}n F\'{\i}sica, Departamento de Ciencias, Pontificia Universidad Cat\'{o}lica del Per\'{u}, Lima, Peru
\item \Idef{org112}SSC IHEP of NRC Kurchatov institute, Protvino, Russia
\item \Idef{org113}Stefan Meyer Institut f\"{u}r Subatomare Physik (SMI), Vienna, Austria
\item \Idef{org114}SUBATECH, IMT Atlantique, Universit\'{e} de Nantes, CNRS-IN2P3, Nantes, France
\item \Idef{org115}Suranaree University of Technology, Nakhon Ratchasima, Thailand
\item \Idef{org116}Technical University of Ko\v{s}ice, Ko\v{s}ice, Slovakia
\item \Idef{org117}Technical University of Split FESB, Split, Croatia
\item \Idef{org118}The Henryk Niewodniczanski Institute of Nuclear Physics, Polish Academy of Sciences, Cracow, Poland
\item \Idef{org119}The University of Texas at Austin, Physics Department, Austin, Texas, United States
\item \Idef{org120}Universidad Aut\'{o}noma de Sinaloa, Culiac\'{a}n, Mexico
\item \Idef{org121}Universidade de S\~{a}o Paulo (USP), S\~{a}o Paulo, Brazil
\item \Idef{org122}Universidade Estadual de Campinas (UNICAMP), Campinas, Brazil
\item \Idef{org123}Universidade Federal do ABC, Santo Andre, Brazil
\item \Idef{org124}University of Houston, Houston, Texas, United States
\item \Idef{org125}University of Jyv\"{a}skyl\"{a}, Jyv\"{a}skyl\"{a}, Finland
\item \Idef{org126}University of Liverpool, Liverpool, United Kingdom
\item \Idef{org127}University of Tennessee, Knoxville, Tennessee, United States
\item \Idef{org128}University of the Witwatersrand, Johannesburg, South Africa
\item \Idef{org129}University of Tokyo, Tokyo, Japan
\item \Idef{org130}University of Tsukuba, Tsukuba, Japan
\item \Idef{org131}Universit\'{e} Clermont Auvergne, CNRS/IN2P3, LPC, Clermont-Ferrand, France
\item \Idef{org132}Universit\'{e} de Lyon, Universit\'{e} Lyon 1, CNRS/IN2P3, IPN-Lyon, Villeurbanne, Lyon, France
\item \Idef{org133}Universit\'{e} de Strasbourg, CNRS, IPHC UMR 7178, F-67000 Strasbourg, France, Strasbourg, France
\item \Idef{org134}Universit\`{a} degli Studi di Pavia, Pavia, Italy
\item \Idef{org135}Universit\`{a} di Brescia, Brescia, Italy
\item \Idef{org136}V.~Fock Institute for Physics, St. Petersburg State University, St. Petersburg, Russia
\item \Idef{org137}Variable Energy Cyclotron Centre, Kolkata, India
\item \Idef{org138}Warsaw University of Technology, Warsaw, Poland
\item \Idef{org139}Wayne State University, Detroit, Michigan, United States
\item \Idef{org140}Wigner Research Centre for Physics, Hungarian Academy of Sciences, Budapest, Hungary
\item \Idef{org141}Yale University, New Haven, Connecticut, United States
\item \Idef{org142}Yonsei University, Seoul, Republic of Korea
\item \Idef{org143}Zentrum f\"{u}r Technologietransfer und Telekommunikation (ZTT), Fachhochschule Worms, Worms, Germany
\end{Authlist}
\endgroup
  %%%%%%% done by webmaster team
\end{document}